\documentclass{emulateapj}

\usepackage{graphicx}

\def\msun{${\rm M}_\odot$}
\def\msunm{{\rm M}_\odot}
\def\lsun{${\rm L}_\odot$}
\def\lsunm{{\rm L}_\odot}
\def\mmax{M_{\rm max}}

\begin{document}

\title{BURST NEUTRINOS FROM NITROGEN FLASH}

\author{A. M. Serenelli}
\affil{Institute for Advanced Study, Einstein Drive, Princeton, NJ 08540, USA}
\and
\author{M. Fukugita}
\affil{Institute for Advanced Study, Einstein Drive, Princeton, NJ 08540, USA}
\affil{Institute  for  Cosmic  Ray  Research, University  of  Tokyo,
  Kashiwa 277-8582, Japan}

\normalsize

\begin{abstract}
Neutrinos   give  a  novel   probe  to   explore  deep   interior  of
astrophysical  objects,  which  otherwise   is  not  accessible  with  optical
observations; among  notable examples are  solar and supernova  neutrinos.  We
show  that there  is  a new  class  of strong  neutrino  emission from  helium
burning, $^{14}{\rm N}+\alpha\rightarrow^{18}{\rm  F}+\gamma$ followed by beta
decay  $^{18}{\rm F}\rightarrow^{18}{\rm O}+e^++\nu_e$,  that gives  a maximum
neutrino luminosity of $10^{8}$ times the solar bolometric luminosity at the 
helium-core flash of  a 1 $M_\odot$ star, whereas the  flash is not observable
by optical means.  This means that the neutrino flux, of average energy of
0.382~MeV, will be 10\% the solar CNO neutrino flux on Earth if the star is
located at 10pc. 
\end{abstract}
 
\maketitle

\section{Preliminaries}\label{sec:intro}

Low mass  stars ($M< 2\msunm$) are  known to develop a  degenerate helium core
during  their evolution  along the  Red  Giant Branch (see for  instance
\citealt{swe78}).  The core
is fed by the surrounding hydrogen-burning shell that slowly moves outwards in
mass as hydrogen  burning proceeds through the CNO  bi-cycle.  The helium mass
fraction  in the hydrogen-depleted  core is  $Y=1-Z$, where  $Z$ is  the total
heavy element  mass fraction  (note that $Z$  does not change  during hydrogen
burning).  As  a result of the  CNO burning, $Z$ is  composed by approximately
70\%  of  $^{14}$N.   When the  mass  of  the  degenerate core  reaches  about
0.47\msun,  the temperature  rises  up  to $10^8$K  and  helium ignites  under
degenerate conditions.  At the
moment of helium ignition, the total  number of $^{14}$N nuclei in the core is
$n(^{14}{\rm N}) \approx  5.3\times10^{53} \left( M_{\rm c}/0.47\right) \left(
Z/0.019 \right)$, where $M_{\rm c}$ is  the core mass in solar mass units.  At
this temperature destruction of $^{14}$N during helium burning (independently
of  the  stellar mass)  occurs  through  the  reaction ${\rm^{14}N}  +  \alpha
\rightarrow   {\rm^{18}F}  +\gamma$  followed   by  beta   decay  ${\rm^{18}F}
\rightarrow {\rm^{18}O}+ e^+  + \nu_e$. The maximum energy  of these neutrinos
is 0.633~MeV, while the average  energy 0.382~MeV.  This implies that were all
the $^{14}$N burned  as described above, a total  energy $\epsilon_\nu \approx
3.2\times10^{47}\left( M_{\rm c}/0.47\right) \left( Z/0.019 \right)$~erg would
be carried away  from the star by neutrinos.  The  importance of this neutrino
emission  contrasts with  the  folklore that  neutrino  emission from  nuclear
reactions 
is not important  during the helium burning phase.  It is  the purpose of this
work to show,  using stellar evolutionary calculations, that  most of this 
energy 
is  actually released  in the  very  short time-scale  that characterizes  the
helium-core flash and that the neutrino flux reaches values as high as
$6\times10^{47}$~s$^{-1}$.

\section{The Helium-core Flash}

The detailed  structure of the degenerate cores  of low mass RGB  stars at the
moment  of  helium  ignition  has  been widely  discussed  in  the  literature
\citep{swe78,swe89,sal02}.  Although  uncertainties  in  the  modeling
exist, a very good understanding of RGB stars has been achieved. This is 
reflected, for instance, in the work by \citet{raf90} where it has been shown 
that observations of globular cluster stars and RR
Lyrae stars indicate that the core mass at helium-flash ignition inferred from
observations and those predicted by stellar evolution calculations agree 
at about a 5\% level. This has been used to constrain different
properties of neutrinos and also on some more exotic particles like axions
\citep{rw92,rw95}.

In  this work,  however, we  are particularly  interested in  nuclear neutrino
production during the helium-core flash, which is a less well understood
evolutionary  phase.   In  particular,   the  central  question  that  remains
to be answered  in a  satisfactorily way  is whether  the helium-core  flash 
 is a
dynamic event  and whether  dynamical episodes are  of any consequence  to the
flash itself and the later evolution.  However, hydrodynamic 2-D and 3-D
calculations (\citealp{deu96,ach95}; and Achatz, M\"uller \& Weiss, in
preparation) show
that hydrostatic  stellar evolution calculations  appear to follow  the global
properties of  the helium-core flash reasonably  well, at least  in its global
properties.  With the confidence
gained from  those results,  we use  for the present  work the  LPCODE stellar
evolution 
code \citep{alt03}. 
Here we review some important  aspects of particular relevance for the present
work.  The nuclear network has been update and now includes  33 isotopes 
from H  to $^{32}$P  and 99 reactions (charged particles reaction rates 
  are taken from \citealp{nacre}). 
Mixing in convective  regions is treated as a diffusive
process and equations are fully coupled to nuclear burning. 
 Conductive
opacities are taken from \citet{hl69}. Neutrino emission rates are  from 
Itoh and collaborators, exception  made of the plasma neutrino emission rate 
that has been adopted from \citet{haft94}. More details can be found in 
\citet{alt03}. 

In what follows, we present the results for the 
evolution of a $1~\msunm$ stellar model of initial solar 
composition $(Y,Z)=(0.275,0.019)$, our reference calculation.

Let us first briefly discuss the  evolution of the model up to the helium-core
flash.  Calculations  were started from  a fully convective  pre-main sequence
model.  The main  sequence  phase lasts  about  9.5~Gyr and  when hydrogen  is
depleted in the  core the star evolves to the RGB  phase, where the degenerate
helium-core grows  as burning in the  hydrogen shell proceeds. As  the mass of
the helium core increases, it contracts and releases gravitational energy that
heats  it up. Additionally,  the temperature  of the  burning shell  is mainly
determined  by  the  local  value  of the  gravitational  potential  and  thus
increases steadily,  also contributing to  increase the core  temperature. The
dominating cooling mechanism in the core is the plasma neutrino emission, that
also increases along with the core size. The net result
is that the  core temperature increases steadily along  the RGB evolution, but
neutrino emission is enhanced closer to the center in the almost isothermal 
core, eventually leading to a temperature inversion.  In our
evolutionary sequence,
this  happens  when the  degenerate  core has  $0.27\msunm$  and  the age  and
luminosity   of   the  model   are   12.62~Gyr   and  $\log(L/{\rm
  L_\odot})$=1.8~ respectively. The model
continues to  evolve along  the RGB with  its luminosity and  core temperature
increasing continuously.  The core degeneracy also continues to increase, 
decoupling  the  pressure  from  the  temperature. This  continues  until  the
temperature  in the  core  is finally  enough  to ignite  the  helium and  the
evolution along  the RGB is halted.  This defines the  tip of the RGB.  In our
calculations, the luminosity at the tip of the RGB for this model 
is $\log(L_{\rm tip}/{\rm L_\odot})$=3.446 and the degenerate helium-core has
0.476~\msun. The age of the  model is 12.664~Gyr. The temperature inversion is
located at $\mmax= 0.23\msunm$; this sets the position where helium ignites. 
At the moment when the  helium-burning luminosity $L_{\rm He}$ exceeds for the
first time that of the hydrogen-burning, the temperature T$_{\rm
  max}$ and density $\rho_{\rm max}$ at $\mmax$ are $1.1\times 10^8$~K and 
$3.2\times10^5$~g~cm$^{-3}$ respectively; the degeneracy
parameter (\citealt{kip90}, Sect. 15.4) is $\psi= 8.5$. 

\begin{figure}
\includegraphics[width=9.cm]{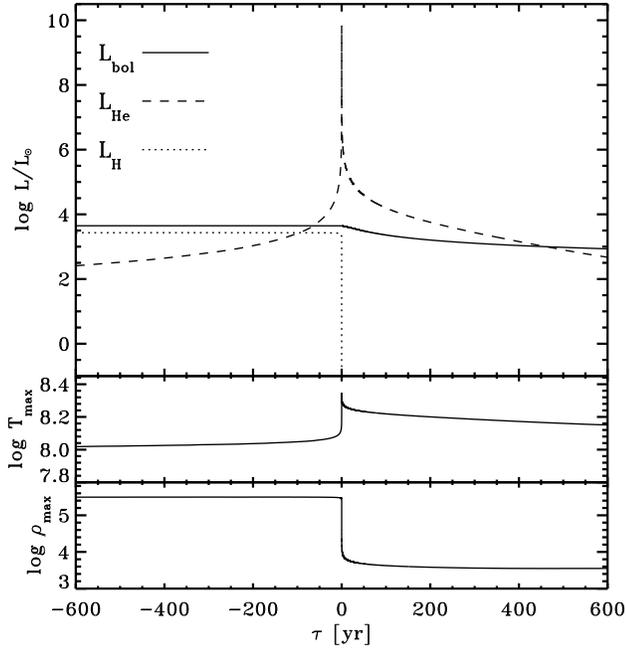}
\figcaption{   Temporal  evolution   of  stellar
quantities  during the  development of  the helium-core  flash in  our 1~\msun
stellar model.   Top panel  shows the evolution  of helium ($L_{\rm  He}$) and
hydrogen ($L_{\rm  H}$) burning luminosities, and  stellar luminosity ($L_{\rm
bol}$) in solar units (\lsun$=3.84\times10^{33}$~erg~s$^{-1}$).  The stellar
luminosity has been shifted up by 0.2 dex to make the otherwise two degenerate
lines ($L_{\rm H}$ and $L_{\rm  bol}$) visible.  Middle and lower panels show,
respectively, the temperature 
and  density at  the point  of maximum  temperature in  the  models ($\mmax$),
located off-center, at about 0.23~\msun. $\tau=0$ corresponds to the moment of
maximum $L_{\rm He}$.
\label{fig:flash}}
\end{figure}

Given the  conditions at the  moment of helium  ignition, the
core is supported by electronic 
degeneracy pressure and, as mentioned before, temperature is decoupled from
pressure.  The
helium flash proceeds through the  triple alpha ($3\alpha$) reaction, that has
a  Maxwellian averaged rate  $N^2_{\rm A}  \left< \sigma  v \right>_{3\alpha}$
very   strongly  dependent   on  temperature:   $N^2_{\rm  A}   \left<  \sigma
v\right>_{3\alpha}   \propto   T^{40,...,20}$   for   $T=   1,...,2.5   \times
10^8$~K. This 
implies that a  small increase in temperature results  in an enormous increase
in the burning  rate, and because the structure of the  star cannot respond to
an increase  in $T$ by expanding  and cooling, a  thermonuclear runaway, the
helium-core flash, quickly follows helium ignition.  
Time evolution  during the  helium-core flash of  some important  quantities is
shown in Fig.~\ref{fig:flash}.  
The top panel presents the time evolution of helium burning
luminosity $L_{\rm He}$. In just a few years, $L_{\rm He}$ grows from
being a negligible energy source in the star to overcome the luminosity of the
star ($L_{\rm bol}$)  by almost seven orders of magnitude.  It is important to
note that when $L_{\rm He}$  reaches about $10^2{\rm L_\odot}$, energy cannot
be transported  by radiation  alone, and adiabatic  convection sets in  in the
layers above 
$\mmax$. At its maximum extension, the convective shell ranges from $\mmax$ up
to $0.475\msunm$.  As shown in the next section, the extension of this
convective shell  determines to  a large extent  the total amount  of N$^{14}$
that is actually burnt during the flash. 
Fig.~\ref{fig:flash}  also shows that  the helium  flash does  not show  up in
$L_{\rm bol}$, which actually
decreases as the hydrogen burning luminosity ($L_{\rm H}$) drops abruptly when
the layers surrounding the core expand and cool. To illustrate the conditions
under  which the  flash starts  and evolves,  the middle  and lower  panels in
Fig.~\ref{fig:flash} show  the temperature and density at  $\mmax$.  The lower
panel, in particular, makes evident that the energy release during the flash 
effectively removes the  degeneracy in the burning region and  is used to lift
up the stellar core from its potential well. 

After the main episode of the helium-core flash, described above, evolutionary
calculations show that a series of
sub-flashes follow, each of them occuring sucessively closer to the center 
until degeneracy is completely removed from the core 
and steady helium-burning is established \citet{ser05}.  
Detailed evolution through this phase is rather uncertain, 
in particular because it is not well understood how heat
flows from the point of maximum temperature inwards (\citealp{ach95,deu96}). 
However, we do not 
expect that  this affects our  main results, because  the bulk of  the nuclear
neutrinos and in  particular the sharp peak in  the neutrino luminosity occurs
during the main flash episode. We show this in the next section.

\section{Neutrinos from the Nitrogen Flash}

We now  focus on  the production of  nuclear neutrinos during  the helium-core
flash. As mentioned before, $^{14}$N is, by far, the second main consituent of
the helium core at the time of helium flash ignition.  The Maxwellian averaged
rate  of the ${\rm  ^{14}N +  \alpha \rightarrow  ^{18}F +  \gamma}$ reaction,
${\rm N_A}  \left< \sigma  v\right>_{14}$ (N$_A$ is  the Avogadro  number), is
between 3 and 5 orders of magnitude larger than that of the $3\alpha$ reaction
in the temperature range characteristic  of the helium flash \citep{nacre} and
its  dependence on  temperature is  extremely high  (${\rm N_A}  \left< \sigma
v\right>_{14}  \propto T^{45,...,21}$  for $T=1,...,2.5  \times  10^8$K).  The
linear dependence  on $\rho$ of the  $^{14} {\rm N} +  \alpha$ reaction delays
its  activation with  respect to  helium ignition  through $3\alpha$,  but the
extreme temperature dependence  leads to a very steep increase  in the rate as
temperature  rises\footnote{The   helium  flash  is,  at   all  times,  driven
by the $3 \alpha$ reaction, while $^{14}$N burning supplies at most
10\% of the nuclear energy release and  for a very short time.  This is so due
to a combination  of factors: when the flash  starts $\rho > 10^5$~g~cm$^{-3}$
and $3\alpha$ dominates because its rate depends as $\rho^2$.  At later times,when  the flash  develops  and starts  producing  n expansion  of the  burning
region, density drops but $^{14}{\rm  N}$ is rapidly depleted.}.  The $^{14}$N
half-life for $\alpha$-captures decreases from $10^{15}$~s when helium ignites
at  $T=10^8$K  to  $1.7\times10^4$~s  at  $T=2.25\times  10^8$K,  the  maximum
temperature reached in the flash.  We call this the ``Nitrogen Flash'', a term
coined by  \citet{iben67} in a  pioneering work on  helium-flash calculations.
At that time,  the $^{14} {\rm N}  + \alpha$ reaction was thought  to halt the
evolution along the Red Giant Branch  and actually lead to the ignition of the
helium-flash. It  seems clear now  that, as described  above, this is  not the
case and $^{14}$N  burning lags slightly behind the  development of the flash,
but we believe the name is still appropriate.

\begin{figure}
\includegraphics[width=9cm]{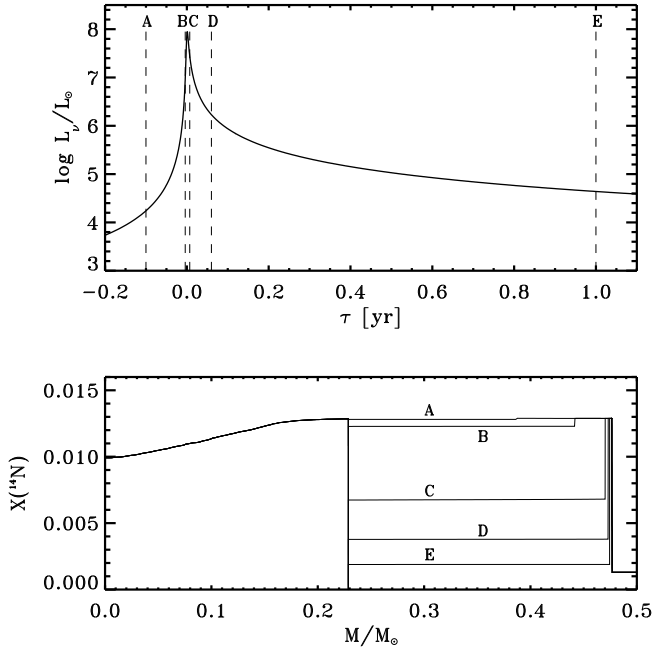}
\figcaption{ Top panel.  The evolution of  the 
nuclear neutrinos  luminosity ($L_\nu$) very  close to the flash  peak ($\tau=
0$). Note that nuclear neutrinos  during the evolutionary phase shown here are
solely created in the beta decay $^{18}{\rm F} \rightarrow + e^+ + \nu_e$ that
follows the $\alpha$-capture $^{14}{\rm N} + \alpha \rightarrow ^{18}{\rm F} 
+\gamma$. Vertical lines,  labeled A, B, C, D and E  denote five chosen epochs
for which the internal abundance profile of $^{14}$N in mass fraction is shown
in the  lower panel.   Initially ({\bf A})  $^{14}$N is barely  depleted.  The
temperature dependence  of the reaction  $^{14}{\rm N}(\alpha,\gamma)^{18}{\rm
F}$ is  even larger than  that of the  $3\alpha$ reaction and  when $^{14}{\rm
  N}(\alpha,\gamma)^{18}{\rm F}$  becomes fully active it produces  a burst of
neutrinos that lasts for less than 4 days 
({\bf B-C}) during which the neutrino luminosity of the star reaches a maximum
value  of  about $10^8\lsunm$  (this  is  about a  5\%  percent  of the  total
luminosity  of nuclear neutrinos  of the  Milky Way).   50\% of  the available
$^{14}$N in the burning region is burnt in this time interval. Strong $^{14}$N
burning continues ({\bf D}) but removal of degeneracy and subsequent expansion
and cooling  of the burning layers quickly  follow and set a  much slower pace
for $^{14}$N depletion ({\bf E}).  The step-like shape of the profiles denotes
the presence  of convection, that keeps  the shell with  an almost homogeneous
composition even when nuclear burning is actually occuring only in the
$0.23-0.27\msunm$  region.  In  passing  by, we  note  that the  shape of  the
$^{14}$N profile in the inner  0.2\msun reflects the increasing temperature of
hydrogen-shell burning with increasing  core size during previous evolutionary
stages.
\label{fig:neutri}}
\end{figure}

The  complete chain  for  $^{14}$N burning  is ${\rm^{14}  N+\alpha\rightarrow
^{18}F}+\gamma$        followed        by        beta       decay        ${\rm
^{18}F\rightarrow^{18}O}+e^++\nu_e$, where  the neutrinos are  produced with a
maximum energy  of 0.633 MeV.   The beta decay  half-life of $^{18}$F  is only
$6.6\times   10^3$~s,   shorter   than   the   half-life   of   $^{14}$N   for
$\alpha$-captures.   Other destruction channels  of $^{18}$F  are at  most two
orders of magnitude slower than the beta decay and do not play any significant
role    in    the    $^{18}$F    nucleosynthesis    (in    particular    ${\rm
^{18}F}(n,\alpha){\rm  ^{15}N}$ dominates  among these  channels when  a small
mass fraction  of neutrons  of about $10^{-19}$  builds up transiently  at the
base  of the  burning  shell).  This  has  the consequence  that the  neutrino
production rate basically follows that  of $^{14}$N destruction.  In the upper
panel  of Fig.~\ref{fig:neutri}  we show  the  time evolution  of the  nuclear
neutrino  luminosity  produced  by   $^{18}$F  decay.   Note  the  very  short
time-scale  that dominates  its evolution,  more  than 4  orders of  magnitude
increase  in only  0.2~years.  The  maximum luminosity  is  about $10^8\lsunm=
3.8\times10^{41}$~erg~s$^{-1}$ and  the resulting number flux  of neutrinos at
the peak is about $6\times 10^{47}$~s$^{-1}$.
We have  shown five epochs of  the internal distribution of  the $^{14}$N mass
fraction in  the lower panel  of Fig.~\ref{fig:neutri} in order  to illustrate
how    $^{14}$N    depletion    and,    accordingly,    neutrino    production
proceeds. $^{14}$N in the shell of $M=0.23$ to 0.475~\msun is destroyed nearly
entirely within 1 yr.  Indeed, $2.65\times10^{-3}$~\msun of $^{14}$N, which is
50\% of  $^{14}$N produced  in the  previous CNO burning  is depleted  in this
short  time-scale.  The  nuclear  neutrino energy  emitted  in nitrogen  flash
amounts to 
0.05\% the value emitted in the entire life of the star.  $^{18}$O produced in
the  beta decay is  eventually all  consumed to  produce $^{22}$Ne  by further
$\alpha$-captures,   which   has   a   reaction   rate   comparable   to   the
$^{14}$N$+\alpha$ reaction,  and therefore it entirely disappears  in the core
of white dwarfs.   The only in-principle visible event  that signatures helium
flash are the burst neutrinos.

The  peak  luminosity  of  nuclear  neutrinos  during  the  helium-core  flash
$(L_\nu^{\rm   peak})$  is   directly  linked   to  the   strength   of  the
helium-flash and thus uncertainties in the helium-flash calculations
will  also affect  the  value of  $(L_\nu^{\rm  peak})$. A  number of  factors
contribute to the uncertainty in the helium flash calculations. 
On one hand, uncertainties
in the input physics is amply dominated by uncertainties of a factor of 2
in the  electron conductive opacities ($\kappa_{\rm cond}$)  in degenerate
cores of RGB stars \citep{sal02,cat05}. 

Other possible sources of uncertainty like 
thermal neutrino emission,  equation of state or radiative  opacities are much
better constrained and have a much  smaller impact on the uncertainties of RGB
stellar  models  \citep{sal02};  and  consequently  in  the  nuclear  neutrino
emission. A second source of uncertainty, that cannot currently be quatified,
is that  introduced by the lack  of complete hydrodynamic  calculations of the
helium-core flash.  Although these  calculations seems to support hydrodynamic
stellar 
evolutionary calculations (in particular  \citealp{ach95} shows that no mixing
beyond  the canonical  convective  boundaries as  given  by the  Schwarzschild
criterion occur and also that the temperature gradient in the convective shell
is very close to the adiabatic one), uncertainties in how heat flows below the
temperature inversion are present and can have some influence on the temporal 
evolution  of the  temperature at  $\mmax$.  Finally, the  peak luminosity  of
nuclear neutrinos depends on the mass and composition of the star. 
We  have  addressed the  first  and  last points  by  performing  a series  of
additional evolutionary calculations.  We do not attempt to cover
the  parameter space, but  just to have  a rough estimation  of the
level of the uncertainties that can be expected. The results are summarized in
Table~\ref{tab:uncert}.  The  behavior of $L_\nu^{\rm  peak}$ at peak  on mass
and 
helium content  reflects that  of $L_{\rm  He}$ at peak.  The behavior  on the
metallicity follows approximately the linear relation expected from the simple
arguments presented  in Sect.~\ref{sec:intro}.  The  dependence of $L_\nu^{\rm
  peak}$  on  $\kappa_{\rm cond}$  seems  to  be linear,  and  a  factor of  2
uncertainty in $\kappa_{\rm cond}$ translates  into a factor of 2 unceratainty
in $L_\nu^{\rm peak}$ at peak.  
A last source of uncertainty, affecting only the neutrino production, is the 
N$^{14}+\alpha$ reaction rate.  Its uncertainty directly translates into an
uncertainty in $L_\nu^{\rm peak}$. 
\citet{gor00} have  determined this reaction rate  to be 2 to  3 times smaller
than  that  of \citealp{nacre}  (the  value adopted  in  this  work), with  an
uncertainty of about 20\% in the temperature range of interest.

\section{Concluding Remarks}

We have shown that during the main episode of the helium-core flash a burst of
neutrinos  is produced  due  to the  extremely  rapid burning  of N$^{14}$  by
$\alpha$ captures in the  so-called Nitrogen Flash. Our reference calculation,
a $1~\msunm$ stellar model  of solar composition $(Y,Z)=(0.275,0.019)$ gives a
nuclear neutrino peak luminosity of about $10^8\lsunm$. 
Neutrinos  from  the  Nitrogen  Flash  carry precious  information  about  the
degenerate  conditions  under  which  the  helium-core flash  is  believed  to
occur. They may be the only direct manifestation we can ever detect of such an
event, that does not manifest itself in the stellar surface. 
The neutrino flux expected  at the flash is $5.3\times10^7(d/10{\rm pc})^{-2}$
cm$^{-2}$s$^{-1}$ at maximum (for a star located $d$ pc away), which lasts for
3 days. This is about 10\% of the CNO neutrino flux from the Sun. 

We are not optimistic about the detection of these burst neutrinos,
not only because this is a rare
event,   but   also   because    solar   neutrinos   will   give   a   serious
background.   A very  large  volume  detector with  the  capability of  energy
resolution  with  the threshold  set  just above  the  maximum  energy of  p-p
neutrinos is needed. To give an  idea, in a 1000 ton hydrocarbon scintillator,
we expect  for a helium flash  at 10~pc 10 events  for 3 days, which  is to be
compared to 80 events from solar  CNO neutrinos for the neutrino energy window
of 0.42  to 0.63~Mev (here neutrino  oscillation is ignored).  In practice, we
also expect a background from solar $^7$Be neutrinos for electron scattering.

\acknowledgements
We want to acknowledge James Gunn and Bohdan Paczy\'nski for valuable
discussions  and comments, John  Bahcall and  Achim Weiss  for  a careful
reading  of  the  manusctript, and  Alan  Chen  for  pointing out  the  latest
reference on the N$^{14}+\alpha$ reaction  rate. We also welcome comments from
the referee, that have helped to improve the original manuscript substantially.
A.M.S.  is supported  in part by the NSF through the  grant PHY-0070928 to the
Institute  for  Advanced Study  and  M.F.  in  part  by  Monell Foundation  at
Princeton.

\newpage

\newpage

\newpage

\begin{table}
\caption{Important quantities characterizing the helium-core flash of  our
  evolutionary sequences.
 \label{tab:uncert} }
\begin{center}
\begin{tabular}{lcccc}
\hline\hline
Model & M$_{c}$ & $\log{L_{\rm Tip}}$ & $\log{L_{\rm He}}$
& $\log{L_\nu}$ \\
\hline
Reference & 0.476 & 3.446 & 9.84 & 7.95 \\
$0.5 \times \kappa_{\rm cond}$ & 0.484 & 3.494 & 10.18 & 8.29 \\
$2 \times \kappa_{\rm cond}$ & 0.468 & 3.403 & 9.58 & 7.65 \\
Y= 0.2505 & 0.480 & 3.440 & 9.97 & 8.12 \\ 
Y= 0.3005 & 0.472 & 3.430 & 9.71 & 7.78 \\ 
Z= 0.029 & 0.474 & 3.442 & 9.78 & 8.07 \\ 
Z= 0.009 & 0.479 & 3.413 & 9.90 & 7.70 \\
Z= 0.0019 & 0.482 & 3.347 & 10.02 & 7.14 \\
M= 0.80 & 0.478 & 3.432 & 9.88 & 8.01 \\ 
M= 1.20 & 0.476 & 3.434 & 9.83 & 7.94 \\ 
M= 1.40 & 0.476 & 3.434 & 9.86 & 7.95 \\
\hline
\end{tabular}
\end{center}
\tablecomments{The  table  summarizes  the  most relevant  quantities  of  our
  evolutionary  sequences.  First  column  denotes  the model  by  giving  the
  parameter that has been modified  with respect to the reference model (first
  row). Second and  third columns give the helium-core  size and luminosity at
  the  RGB-tip. Fourth  and fifth  columns  give the  peak helium-burning  and
  nuclear neutrino  luminosities during the  flash. In all  cases luminosities
  are in solar units.}  
\end{table}


\newpage

\thebibliography{99}


\bibitem[Achatz(1995)]{ach95}
Achatz, K. 1995, Master's Thesis, Tech. Univ. Munich

\bibitem[Angulo et~al.(1999)]{nacre}
Angulo, C. et al. 1999, Nuc. Phys. A, 656, 3

\bibitem[Althaus et al.(2003)]{alt03}
Althaus, L.~G., Serenelli, A.~M., C\'orsico, A.~H., \& Montgomery, M.~H. 2003,
A\&A, 404, 593

\bibitem[Catelan(2005)]{cat05}
Catelan, M.  2005, in ASP Conf. Ser. TBA, Resolved Stellar Populations, ed. 
D. Valls-Gabaud, \& M. Chavez, to be published, preprint 
(astro-ph/0507464)

\bibitem[Deupree(1996)]{deu96}
Deupree, R. G. 1996, \apj, 471, 377

\bibitem[G\"orres et al.(2000)]{gor00}
G\"orres, J., Arlandini, C., Giesen, U., Heil, M., Käppeler, F., Leiste, H.,
Stech, E., \& Wiescher, M. 2000, \prc, 62, 055801

\bibitem[Haft et al.(1994)]{haft94}
Haft, M., Raffelt, G., \& Weiss, A. 1994, \apj, 425, 222

\bibitem[Hubbard \& Lampard(1969)]{hl69}
Hubbard, W. B., \& Lampe, M. 1969, \apjs, 19, 297

\bibitem[Iben(1967)]{iben67}
Iben, I., Jr. 1967, \apj, 147, 650

\bibitem[Itoh et al.(1983)]{itoh83}
Itoh, N., Mitake, S., Iyetomi, H., \& Ichimaru, S. 1983, \apj, 273, 774



\bibitem[Kippenhahn \& Weigert(1990)]{kip90}
Kippenhahn, R. \& Weigert A. 1990, in {\it Stellar Structure and Evolution}, 
Springer-Verlag, Berlin 

\bibitem[Raffelt(1990)]{raf90}
Raffelt, G. G. 1990, \apj, 365, 559

\bibitem[Raffelt \& Weiss(1992)]{rw92}
Raffelt, G. G., \& Weiss, A. 1992, \aap, 264, 536

\bibitem[Raffelt \& Weiss(1995)]{rw95}
Raffelt, G. G., \& Weiss, A. 1995, \prd, 51, 1495

\bibitem[Salaris et al.(2002)]{sal02}
Salaris, M., Cassisi, S., \& Weiss, A. 2002, PASP, 114, 375

\bibitem[Schwarzschild \& H\"arm(1962)]{sch62}
Schwarzschild, M., \& H\"arm, R. 1962, ApJ, 136, 158

\bibitem[Serenelli \& Weiss(2005)]{ser05}
Serenelli,  A.   M., \&  Weiss,  A.   2005,  \aap, accepted  for  publication,
preprint (astro-ph/0507430)

\bibitem[Sweigart et al.(1989)]{swe89}
Sweigart, A.~V., Greggio, L., \& Renzini, A. 1989, \apjs, 69, 911

\bibitem[Sweigart \& Gross(1978)]{swe78}
Sweigart, A.~V., \& Gross, P.~G. 1978, ApJ, 36, 405
\end{document}